\documentclass{jfm}

\usepackage{subfigure}
\usepackage{graphicx}
\usepackage{natbib}
\usepackage{bm}

\ifCUPmtlplainloaded \else
  \checkfont{eurm10}
  \iffontfound
    \IfFileExists{upmath.sty}
      {\typeout{^^JFound AMS Euler Roman fonts on the system,
                   using the 'upmath' package.^^J}%
       \usepackage{upmath}}
      {\typeout{^^JFound AMS Euler Roman fonts on the system, but you
                   dont seem to have the}%
       \typeout{'upmath' package installed. JFM.cls can take advantage
                 of these fonts,^^Jif you use 'upmath' package.^^J}%
      }
  \else
  \fi
\fi

\ifCUPmtlplainloaded \else
  \checkfont{msam10}
  \iffontfound
    \IfFileExists{amssymb.sty}
      {\typeout{^^JFound AMS Symbol fonts on the system, using the
                'amssymb' package.^^J}%
       \usepackage{amssymb}%
         \let\leq=\leqslant
         \let\geq=\geqslant
      }{}
  \fi
\fi

\ifCUPmtlplainloaded \else
  \IfFileExists{amsbsy.sty}
    {\typeout{^^JFound the 'amsbsy' package on the system, using it.^^J}%
     \usepackage{amsbsy}}
    {\providecommand\boldsymbol[1]{\mbox{\boldmath $##1$}}}
\fi

\newsavebox{\astrutbox}
\sbox{\astrutbox}{\rule[-5pt]{0pt}{20pt}}

\title[Counter-gradient heat transport in 2D turbulent RB convection]{Counter-gradient heat transport in two-dimensional turbulent Rayleigh-B\'{e}nard convection}
\author[Y.-X. Huang $\&$ Q. Zhou]{Yong-Xiang HUANG and Quan ZHOU\thanks{Email address for correspondence: qzhou@shu.edu.cn}}
\affiliation{Shanghai Institute of Applied Mathematics and Mechanics, and Shanghai Key Laboratory of Mechanics in Energy Engineering, Shanghai University, Shanghai 200072, China}

\date{?? and in revised form ??}

\begin{document}

\maketitle

\begin{abstract}

We present high-resolution numerical investigations of heat transport by two-dimensional (2D) turbulent Rayleigh-B\'{e}nard (RB) convection over the Rayleigh number range $10^8 \leqslant Ra\leqslant 10^{10}$ and the Prandtl number range $0.7\leqslant Pr \leqslant10$. We find that there exist strong counter-gradient local heat flux with magnitude much larger than the global Nusselt number $Nu$ of the system. Two mechanisms for generating counter-gradient heat transport are identified: one is due to the bulk dynamics and the other is due to the competitions between the corner-flow rolls and the large-scale circulation (LSC). While the magnitude of the former is found to increase with increasing  Prandtl number, that of the latter maximizes at medium $Pr$. We further reveal that the corner-LSC competitions lead to the anomalous $Nu$-$Pr$ relation in 2D RB convection, i.e. $Nu(Pr)$ minimizes, rather than maximizes as in three-dimensional cylindrical case, at $Pr\approx2\sim3$ for moderate $Ra$.

\end{abstract}

\section{Introduction}

The convective motion of enclosed fluids is of fundamental interest, as well as being present in a variety of engineering, geophysical, and astrophysical systems. A paradigmatic model that has been widely used to study this type of flow is Rayleigh-B\'{e}nard (RB) convection (Ahlers, Grossmann $\&$ Lohse 2009; Lohse $\&$ Xia 2010), i.e. a fluid layer heated from below and cooled on the top. A central issue in the study of RB system is to understand how heat is transported upwards across the fluid layer by convective flows \cite[]{agl2009rmp, cs2012epje, ahlers2012prl, xia2013prl, xia2013taml, lohse2013pnas, urban2013pnas}. It is usually measured in terms of the Nusselt number $Nu$ $(=QH/\chi\Delta)$, which depends on the control parameters of the system, such as the Rayleigh number $Ra$ $(=\alpha gH^3\Delta/\nu\kappa)$ and the Prandtl number $Pr$ $(=\nu/\kappa)$. Here, $Q$ is the heat current density across the fluid layer of thermal conductivity $\chi$ with height $H$ and with an applied temperature difference $\Delta$, $g$ is the acceleration due to gravity, and $\alpha$, $\nu$, and $\kappa$ are, respectively, the thermal expansion coefficient, kinematic viscosity, and thermal diffusivity of the working fluid. Specifically, the $Pr$-dependence of $Nu$ has long been studied (Kraichnan 1962; Verzicco $\&$ Camussi 1999; Kerr $\&$ Herring 2000; Ahlers $\&$ Xu 2001; Grossmann $\&$ Lohse 2001; Xia, Lam $\&$ Zhou 2002; Roche \emph{et al.} 2002; Niemela $\&$ Sreenivasan 2003; Breuer \emph{et al.} 2004; Calzavarini \emph{et al.} 2005; Silano, Sreenivasan $\&$ Verzicco 2010; Stevens, Lohse $\&$ Verzicco 2011; Zhou \emph{et al.} 2012, 2013), especially in three-dimensional (3D) cylindrical samples. It is found that for moderate $Ra$, with increasing Prandtl number, $Nu$ first increases, reaches its maximum value at around $Pr\approx3$ (depends on $Ra$), and then slightly decreases \cite[]{ahlers2001prl, gl2001prl, xia2002prl}. For higher $Ra$, $Nu$ seems to become independent of $Pr$ \cite[]{ssv2010jfm, stevens2011jfm}.

In the field of convection, two-dimensional (2D) RB flow has played an important role (Zhang $\&$ Wu 2005; Zhang, Wu $\&$ Xia 2005), partly because its relevance to thermal convection occurring in the atmosphere \cite[]{seychelles2008prl, seychelles2010prl}. Recently, 2D convection was also utilized as a test-bed to study the physical and turbulent transport features of 3D convection, e.g., it was adopted to compare turbulent heat transport between conditions of constant temperature and constant heat flux \cite[]{johnston2009prl}, to reveal the non-Boussinesq effects on the flow organization \cite[]{sugiyama2009jfm}, to analyze the reversals of the large-scale circulation (LSC) \cite[]{xgl2010prl, chandra2011pre, chandra2013prl}, to study the boundary layer (BL) dynamics \cite[]{zx2010jfm,zhou2011pof}, and to connect the flow structures and heat flux (van der Poel, Stevens $\&$ Lohse, 2011; van der Poel \emph{et al.} 2012). Moreover, many heat-transfer theories for turbulent RB system are essentially 2D, e.g. the Grossmann-Lohse theory \cite[]{gl2000jfm} and the Whitehead-Doering theory for the ultimate regime \cite[]{doering2011prl}. In general, 2D geometry offers two advantages: (i) the numerical effort for 2D simulations is much smaller so that a full resolution of the BLs at high Rayleigh numbers can be guaranteed; (ii) the flow visualizations of the full temperature and velocity fields are much easier so that a direct connection between the flow organization and the heat-transfer properties is possible.

Nevertheless, we note that there is still lack of systematic studies on the heat-transfer properties in 2D RB convection, such as the $Pr$-dependence of $Nu$. In this paper, we will fill this gap with the help of direct numerical simulations (DNS).  The simulations were made over the ranges $10^8\leq Ra\leq10^{10}$ and $0.7\leq Pr\leq10$. We show that heat transfer of 2D RB flow for not too large $Ra$ exhibits a very different $Pr$-dependence from the 3D cylindrical case, i.e. $Nu(Pr)$ minimizes, rather than maximizes as in 3D case, around $Pr\approx2\sim3$. We further show that such a behavior is due to counter-gradient heat transport generated by the competitions between the corner-flow rolls and the LSC. The remainder of this paper is organized as follows. $\S$2 introduces the numerical method adopted in the present paper. The $Pr$-dependence of $Nu$ is discussed in $\S$3, where we try to establish a direct connection between the flow structures and the $Nu$-$Pr$ behaviors. We summarize our findings and conclude in $\S$4.

\section{Numerical Method}

We deal with a wall-bounded 2D domain of height $H=1$ and horizontal length $D=1$ with uniform grids. The numerical code is based on a compact fourth-order finite difference scheme of the time-dependent incompressible Oberbeck-Boussinesq equations in vorticity-stream function formulation, i.e.
\begin{equation}
\frac{\partial \omega}{\partial t} + (\mathbf{u}\cdot\nabla)\omega=\nu\nabla^2\omega+\alpha g\frac{\partial\theta}{\partial x},
\label{eq:ns_omega}
\end{equation}
\begin{equation}
\nabla^2\psi=\omega,
\label{eq:psi}
\end{equation}
\begin{equation}
u = -\frac{\partial\psi}{\partial z}, \mbox{\ \ }w=\frac{\partial\psi}{\partial x}.
\label{eq:uw}
\end{equation}
\begin{equation}
\frac{\partial\theta}{\partial t}+(\mathbf{u}\cdot\nabla)\theta=\kappa\nabla^2\theta,
\label{eq:ns_t}
\end{equation}
Here, $\mathbf{u}(x,z,t)=u\vec{x}+w\vec{z}$ is the velocity field ($\vec{x}$ and $\vec{z}$ are the horizontal and vertical unit vectors, respectively), $\omega$ is the vorticity, $\psi$ is the stream function, and $\theta$ is the temperature. The scheme was proposed by Liu, Wang $\&$ Johnston (2003) and the accuracy, stability, and efficiency of the scheme have been examined in detail by \cite{liu1996jcp} and \cite{liu2003jsc}. It has also been shown that the global quantities, such as $Nu$, obtained using the compact scheme agree well with those obtained using other numerical schemes, such as the Fourier-Chebyshev spectral collocation method \cite[]{johnston2009prl}. Recently, we have applied the same numerical code to study small-scale properties in 2D Rayleigh-Taylor turbulence \cite[]{zhou2013pof}. Here we briefly describe the scheme. For equations (\ref{eq:ns_omega})-(\ref{eq:uw}), an essentially compact fourth-order (EC4) scheme, first proposed by \cite{liu1996jcp} for the 2D Navier-Stokes equations, is employed to solve the momentum equations with the gravity term treated explicitly. \cite{liu1996jcp} have shown that the EC4 scheme has very nice features with regard to the treatment of boundary conditions. Such a scheme is also very efficient, because at each Runge-Kutta stage only two Poisson-like equations have to be solved by taking the standard fast fourier transform based Poisson solvers. The heat transfer equation (\ref{eq:ns_t}) is treated as a standard convection-diffusion equation and is discretized using fourth-order long-stencil difference operators. Third-order Runge-Kutta method is employed to integrate the equations in time. The time step is chosen to fulfill the Courant-Friedrichs-Lewy (CFL) conditions, i.e., the CFL number is 0.2 or less for all computations presented in this paper.

The grid resolution has been chosen to achieve a full resolution of the BLs \cite[]{shishkina2010njp} and to resolve the smallest scale of the problem, i.e. the Kolmogorov scale $\eta_K$ and the Batchelor scale $\eta_B$. In the present study, the number of grid points is increased from $1025\times1025$ to $4097\times4097$ as $Ra$ increases from $10^8$ to $10^{10}$. Thus, for all runs the thermal BLs are resolved with at least 18 grid points and the grid spacings $\Delta_g<0.3\eta_K$ and $\Delta_g<0.45\eta_B$. No-penetration and no-slip velocity boundary conditions are applied to all four solid walls, which are recast in terms of $\psi$ as $\psi|_{x=0,1; z=0,1}=0$, $\partial\psi/\partial x|_{x=0,1}=0$, and $\partial\psi/\partial z|_{z=0,1}=0$. For temperature the two vertical sidewalls ($x=0,1$) are chosen to be adiabatic (no flux), while at the colder top ($z=1$) and the warmer bottom ($z=0$) plates, the temperature are fixed at $\theta_{cold}=-0.5$ and $\theta_{hot}=0.5$, respectively, and thus the temperature difference across the fluid layer is $\Delta=\theta_{hot}-\theta_{cold}=1$  and the mean bulk temperature, i.e. the arithmetic mean value of $\theta_{cold}$ and $\theta_{hot}$, is $\theta_{bulk}=0$.

\section{Results and discussion}

\begin{figure}
\begin{center}
\resizebox{0.85\columnwidth}{!}{%
  \includegraphics{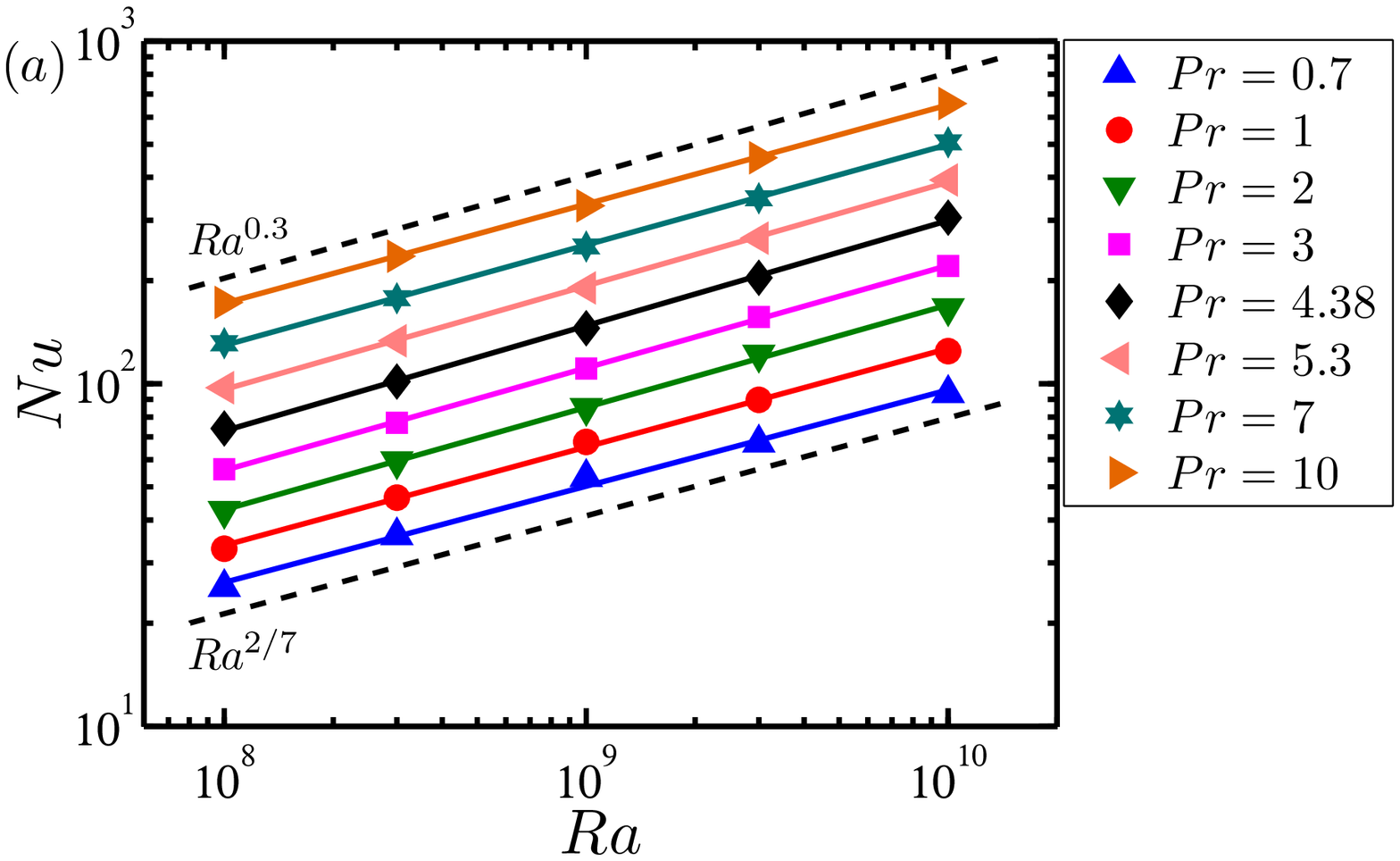}
}
\resizebox{0.495\columnwidth}{!}{%
  \includegraphics{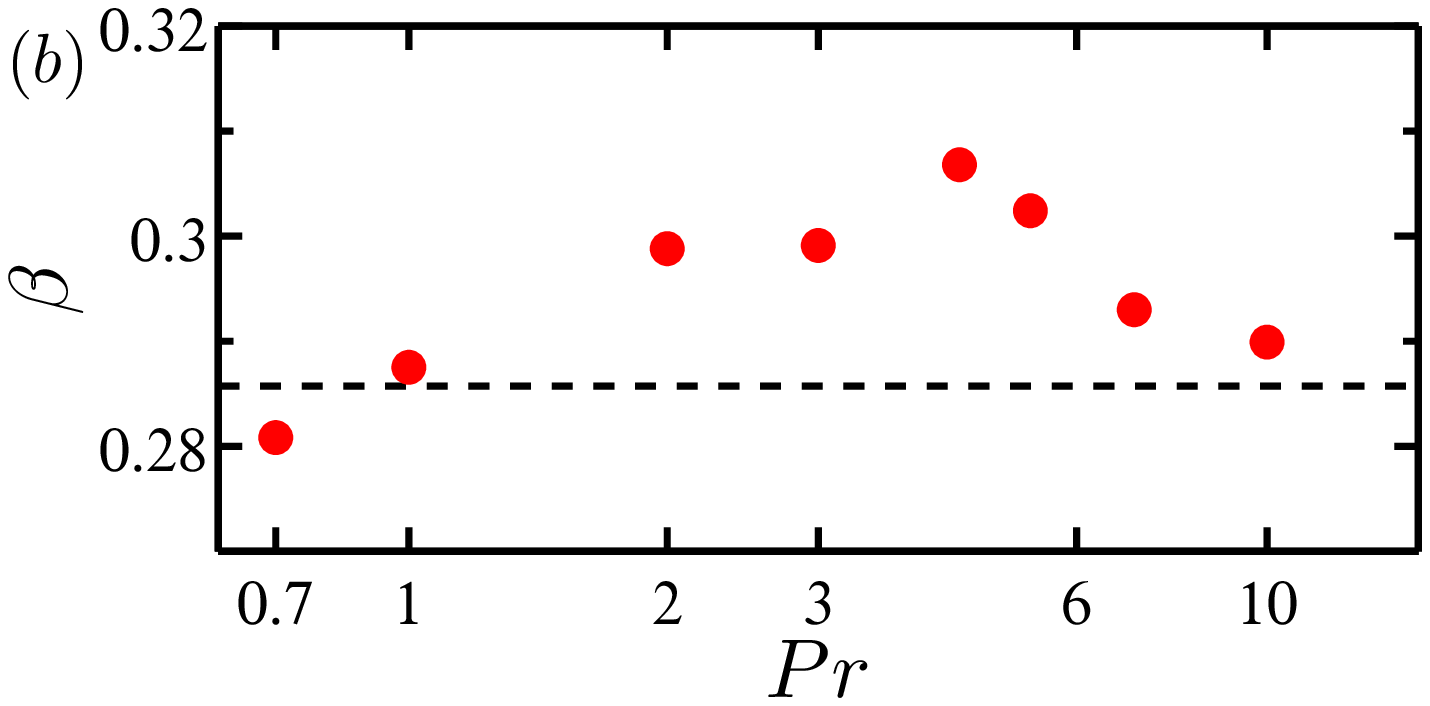}
}
\resizebox{0.495\columnwidth}{!}{%
  \includegraphics{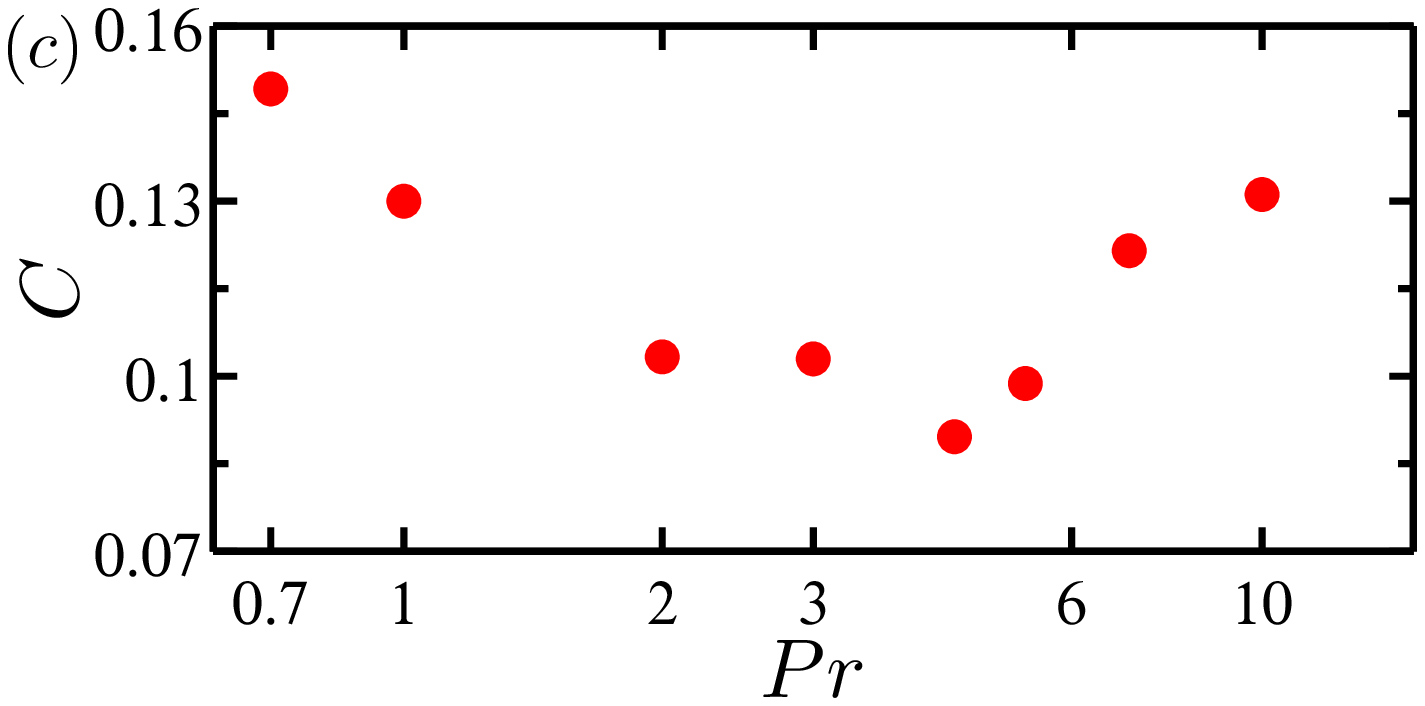}
}
\caption{(Color online). (\emph{a}) Log-log plot of $Nu$ as a function of $Ra$ for 8 values of $Pr$. For clarity, each data set has been shifted upwards from its lower neighbor by a factor of 1.3. The solid lines are the best power-law fits to the corresponding data and the two dashed lines mark the $Ra^{0.3}$ and $Ra^{2/7}$ scalings for reference. (\emph{b}) The fitted scaling exponent $\beta$ of $Nu(Ra)$ as a function of $Pr$. The dashed line indicates the value of 2/7 for reference. (\emph{c}) The fitted prefactor $C$ of $Nu(Pr)$ as a function of $Pr$.} \label{fig:figNuRa}
\end{center}
\end{figure}

Figure \ref{fig:figNuRa} (\emph{a}) shows a log-log plot of the measured $Nu$ as a function of $Ra$ for eight values of $Pr$. For clarity, the data set for each $Pr$ has been shifted upwards from its lower neighbor by a factor of 1.3. Here, $Nu$ is calculated over the whole volume and over a very large averaging time, e.g. more than 1000$t_E$ (400$t_E$) for $Ra=10^8$ ($10^{10}$), where $t_E=4\pi/\langle|\omega_c(t)|\rangle$ is the large eddy turnover times with $\omega_c$ being the center vorticity. The time convergence for the obtained $Nu$ is checked by comparing the time-averages over the first and the last halves of the simulation and the resulting convergence is for most cases smaller than $1\%$. In the figure, we note that the $Nu$-$Ra$ data for each $Pr$ could be well described by the power-law relation, $Nu=C(Pr)Ra^{\beta(Pr)}$ (see the solid lines in the figure). The best-fitted scaling exponent $\beta$ and prefactor $C$ as functions of $Pr$ are plotted in figure \ref{fig:figNuRa} (\emph{b}) and (\emph{c}), respectively, and one can see that both $\beta$ and $C$ vary slightly with $Pr$. We further notice that the obtained scaling exponents $\beta$ are in general agreement with those found in 3D convection cells \cite[]{agl2009rmp, cs2012epje}, while the values of $C$ are smaller than their 3D counterparts.

\begin{figure}
\begin{center}
\resizebox{0.8\columnwidth}{!}{%
  \includegraphics{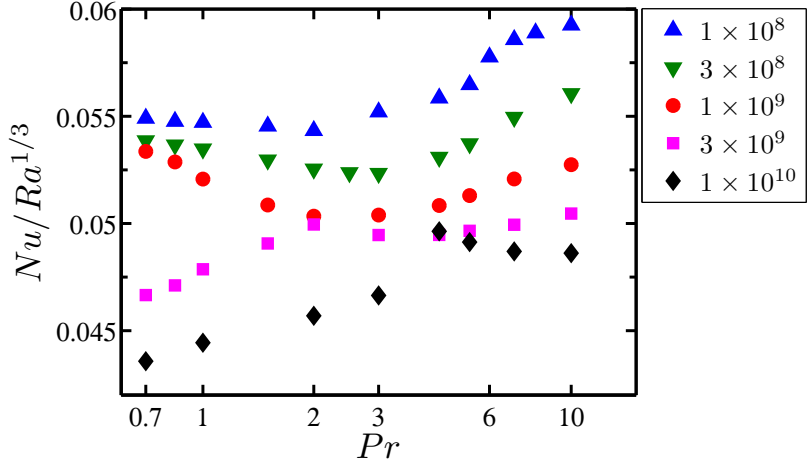}
}
\caption{(Color online). Semilog plot of $Nu/Ra^{1/3}$ versus $Pr$ for $Ra=1\times10^8$, $3\times10^8$, $1\times10^9$, $3\times10^9$, and $1\times10^{10}$.} \label{fig:fig1}
\end{center}
\end{figure}

In figure \ref{fig:fig1} we show the compensated $Nu/Ra^{1/3}$ as a function of $Pr$ for five values of $Ra$. It is seen that for $Ra\leq10^9$, with increasing $Pr$, $Nu$ first decreases, reaches its minimum value at $Pr\approx2\sim3$, and then increases. This is dramatically different from the 3D cylindrical case, where $Nu$ is found to maximize at $Pr\approx3$ for moderate $Ra$ \cite[]{ahlers2001prl, xia2002prl, stevens2011jfm}. It is further seen that the effect of the $Pr$ number on the heat transfer reduces with increasing $Ra$, i.e., the variations in $Nu(Pr)$, as defined by the difference between the maximum and minimum in the Nusselt number series, yield $9.1\%$, $7.1\%$, and $6.0\%$ for $Ra=10^8$, $3\times10^8$, and $10^9$, respectively. For $Ra\geq3\times10^9$ the observed $Pr$-dependence is qualitatively similar to that of the 3D cylindrical case.

\begin{figure}
\begin{center}
\resizebox{0.7\columnwidth}{!}{%
  \includegraphics{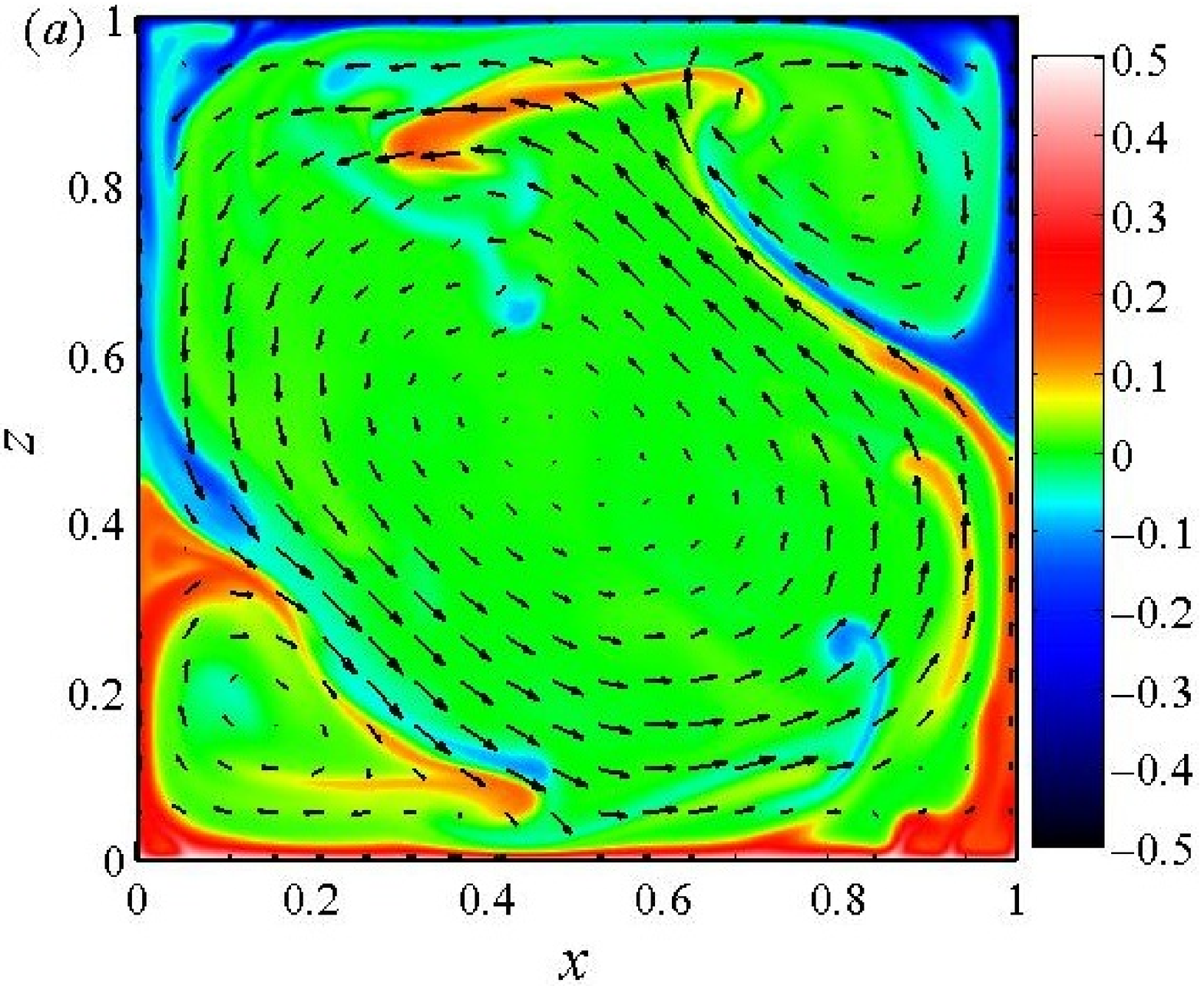}
}
\resizebox{0.7\columnwidth}{!}{%
  \includegraphics{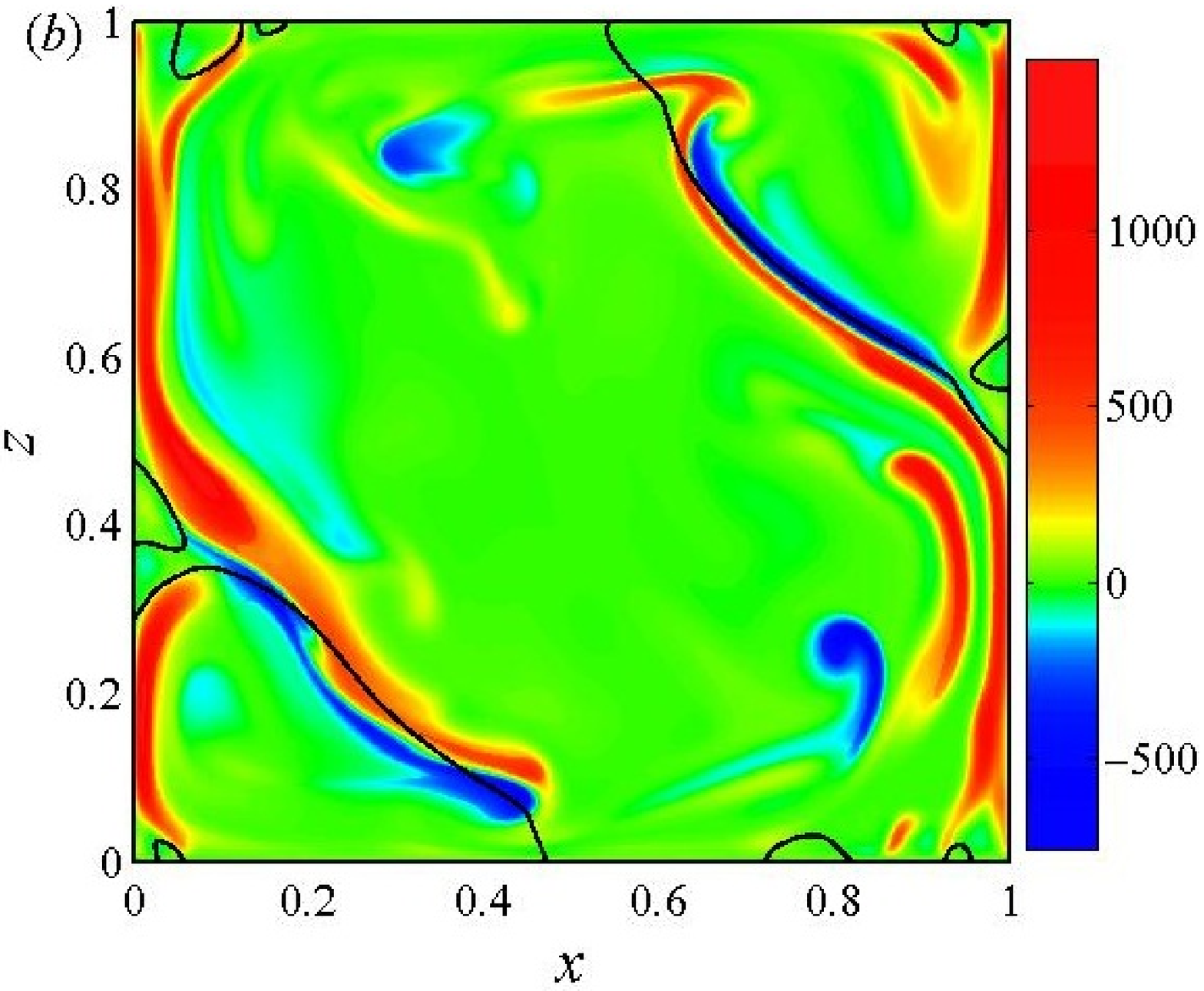}
}
\caption{(Color online). (\emph{a}) Typical snapshot of the instantaneous temperature (color) and velocity (arrows) fields for $Ra=3\times10^8$ and $Pr=4.38$. (\emph{b}) The corresponding snapshot of the local heat flux field (color). The black solid lines mark the streamlines of $\psi=0$, which can roughly distinguish the regions of the corner-flow rolls and the LSC.} \label{fig:fig2}
\end{center}
\end{figure}

To understand the anomalous $Nu$-$Pr$ relation in 2D case, we then study the instantaneous flow structures and try to reveal the connection between the heat-transfer properties and the flow organization. Figure \ref{fig:fig2}(\emph{a}) shows a typical snapshot of the temperature and velocity fields. Corresponding video can be viewed as Supplementary Material. One can see clearly that the dominant flow pattern is a counter-clockwise rotating roll orientated diagonally in the cell. There are still several smaller secondary rolls at the corners of the cell: two larger clockwise rotating rolls at the lower-left and upper-right corners. As pointed out by \cite{xgl2010prl}, the corner rolls are energetically fed by thermal plumes detaching from the BLs. Figure \ref{fig:fig2}(\emph{b}) shows the corresponding snapshot of the local heat flux field, $Nu_L=H(w\theta/\kappa-\partial\theta/\partial z) / \Delta$. It is surprising that there exist strong negative (i.e. opposite to the temperature gradient of the system) $Nu_L$ with magnitude much larger than the global heat transport of the system ($Nu=35.5$ for $Ra=3\times10^8$ and $Pr=4.38$). From the movie corresponding to figure \ref{fig:fig2} two mechanisms for generating counter-gradient local heat transport can be identified: (i) Hot plumes detach from the bottom thermal BL and move upwards with the LSC. After reaching the other plate, some plumes do not lost their thermal energy completely, i.e. they are still hotter than the environment. These plumes continue moving with the LSC, leading to the sinking of the hot fluids and thus generating negative $Nu_L$. The same process can be observed for cold plumes detached from the top plate. This kind of counter-gradient heat transport is due to the bulk dynamics and was also observed in previous 3D RB experiments \cite[]{shang2003prl, shang2004pre, ching2004prl, pinton2007prl}. (ii) When hot plumes move upwards with the lower-left corner rolls, they encounter with the LSC. On the one hand, the corner rolls grow in kinetic energy and thus also in size due to the energy gained from detaching plumes from thermal BLs. On the other hand, the growth of the sizes of the corner rolls is suppressed by the LSC. Therefore, the corner-flow rolls and the LSC compete with each other. Because the strength of the corner rolls is smaller than that of the LSC, these hot plumes are forced to fall back and thus leading to the negative $Nu_L$. The same process can also be observed for cold plumes in the upper-right corner rolls. This kind of counter-gradient heat transport is due to the competitions between the corner rolls and the LSC. Taken together, therefore, both the LSC and the corner flows sometines contribute to heat transport in the `wrong direction': cold fluids can be brought back to the cold plate and hot fluids can be brought back to the hot plate by either the corner flows or the LSC.

\begin{figure}
\begin{center}
\resizebox{0.7\columnwidth}{!}{%
  \includegraphics{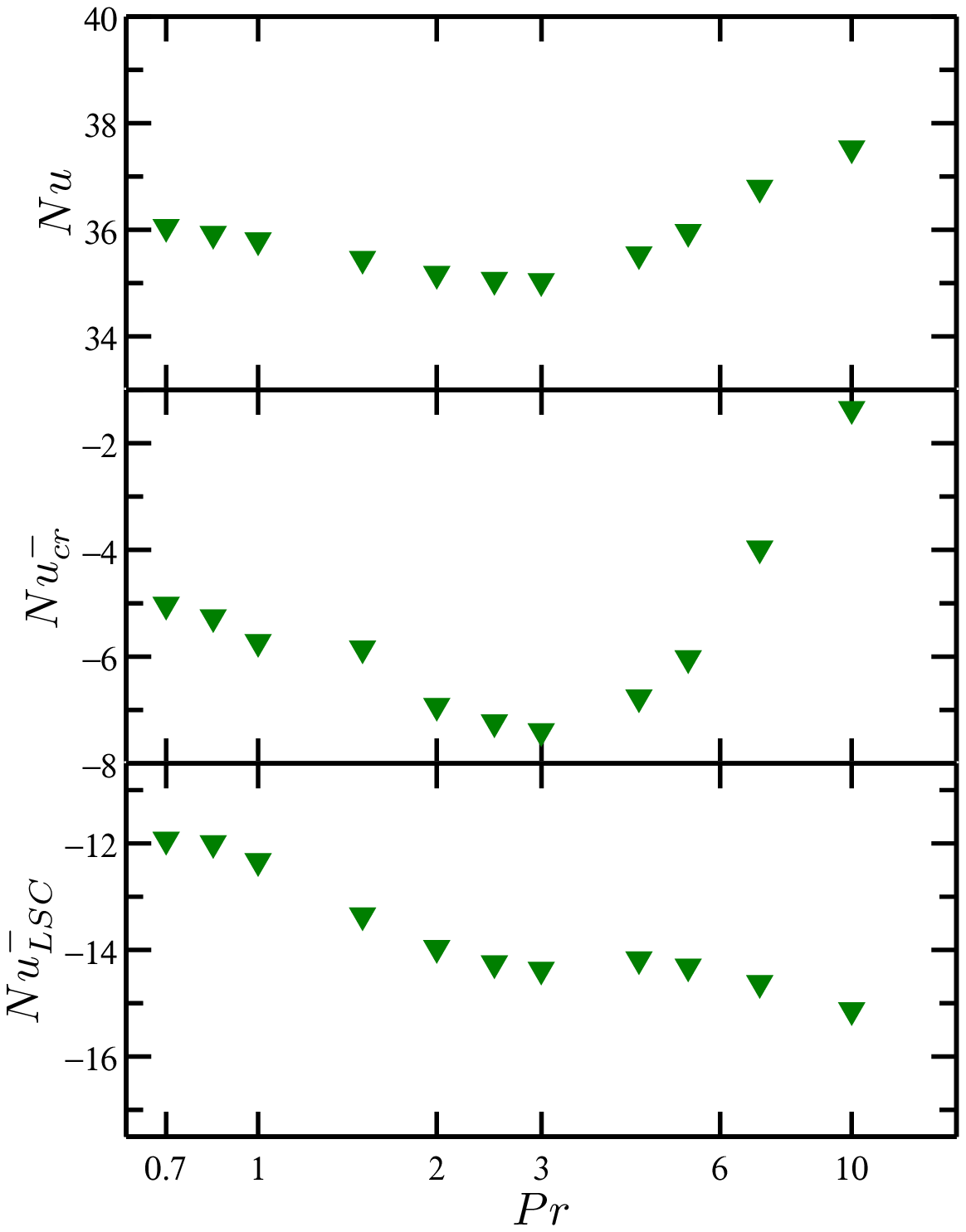}
}
\caption{(Color online). $Pr$-dependence of $Nu$, $Nu^-_{cr}$, and $Nu^-_{LSC}$ for $Ra=3\times10^8$ in semilog plots.} \label{fig:fig3}
\end{center}
\end{figure}

To quantify these two kinds of negative $Nu_L$, we note that the steamlines of the stream function $\psi=0$ can roughly distinguish the regions of the corner rolls and the LSC [see solid lines in figure \ref{fig:fig2}(\emph{b})]. Then the counter-gradient heat flux due to the bulk dynamics, $Nu^-_{LSC}$, and that due to the corner-LSC competitions, $Nu^-_{cr}$, are calculated via conditional average as
\begin{equation}
Nu^-_{LSC}=\langle Nu_L|Nu_L<0\mbox{\ \ }\&\mbox{\ \ }\psi<0\rangle
\end{equation}
and
\begin{equation}
Nu^-_{cr}=\langle Nu_L|Nu_L<0\mbox{\ \ }\&\mbox{\ \ }\psi>0\rangle.
\end{equation}
Figure \ref{fig:fig3} shows the $Pr$-dependence of $Nu$, $Nu^-_{cr}$, and $Nu^-_{LSC}$ for $Ra=3\times10^8$. While the magnitude of $Nu^-_{LSC}$ increases at increasing $Pr$, $Nu^-_{cr}$ is seen to share the similar $Pr$-dependence as the global $Nu$, suggesting, from our point of view, that the observed anomalous 2D $Nu$-$Pr$ relation could be attributed to the corner-LSC competitions. Note that while there is strong negative $Nu_L$ with magnitude much larger than the global $Nu$, the averaged $Nu^-_{LSC}$ and $Nu^-_{cr}$ are smaller in magnitude than the global $Nu$, suggesting this average could be controlled by a few strong events.

The next issue is to physically understand the $Pr$-behaviors of $Nu^-_{cr}$ and $Nu^-_{LSC}$. For $Nu^-_{LSC}$, with decreasing $Pr$ (i.e. increasing thermal diffusivity), plumes become much easier to lost their thermal energy through thermal diffusion, thus resulting in the decrease of the magnitude of $Nu^-_{LSC}$. For $Nu^-_{cr}$, the magnitude of $Nu^-_{cr}$ is related to the strength of the corner flow: stronger corner flows contain more thermal energy, and thus lead to the increase of the magnitude of $Nu^-_{cr}$ after they are forced to move back by the corner-LSC competitions. As argued by \cite{xgl2010prl}, the buildup of the corner flow is suppressed for too small and too large $Pr$. This is because that for too small $Pr$, as discussed above, the thermal energy carried by plumes is lost through thermal diffusion and for too large $Pr$ the thermal coupling between the corner flow and the thermal BL is hindered as the thermal BL is nested in the kinematic BL \cite[]{xgl2010prl}. Therefore, the strength of the corner flow is expected to become strongest and thus the magnitude of $Nu^-_{cr}$ is maximal at some moderate $Pr$, which is in consistent with the results shown in the middle panel of figure \ref{fig:fig3}.

\begin{figure}
\begin{center}
\resizebox{0.495\columnwidth}{!}{%
  \includegraphics{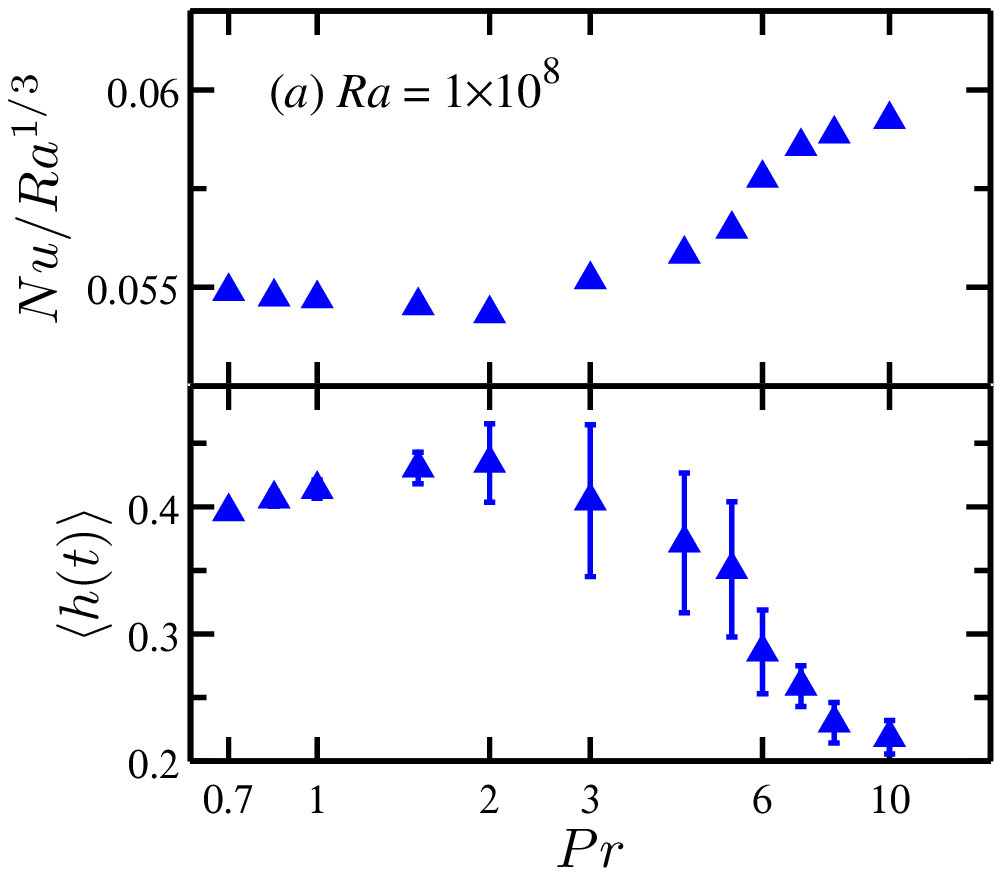}
}
\resizebox{0.495\columnwidth}{!}{%
  \includegraphics{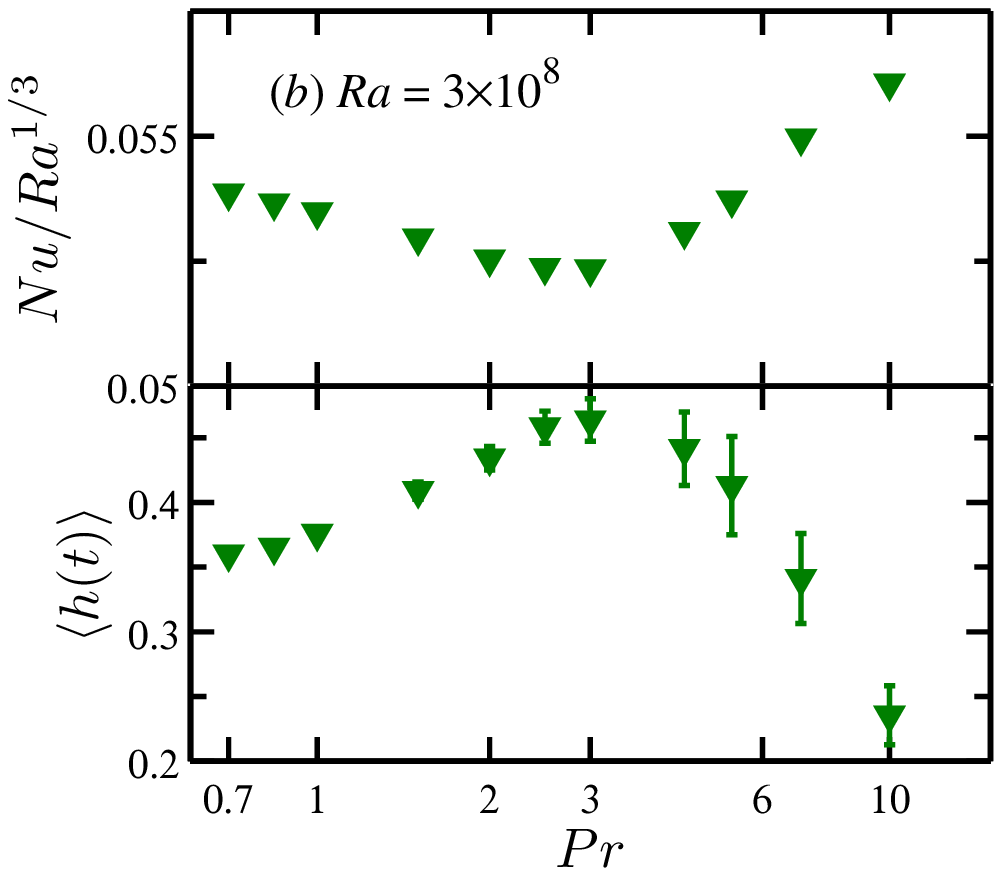}
}
\resizebox{0.495\columnwidth}{!}{%
  \includegraphics{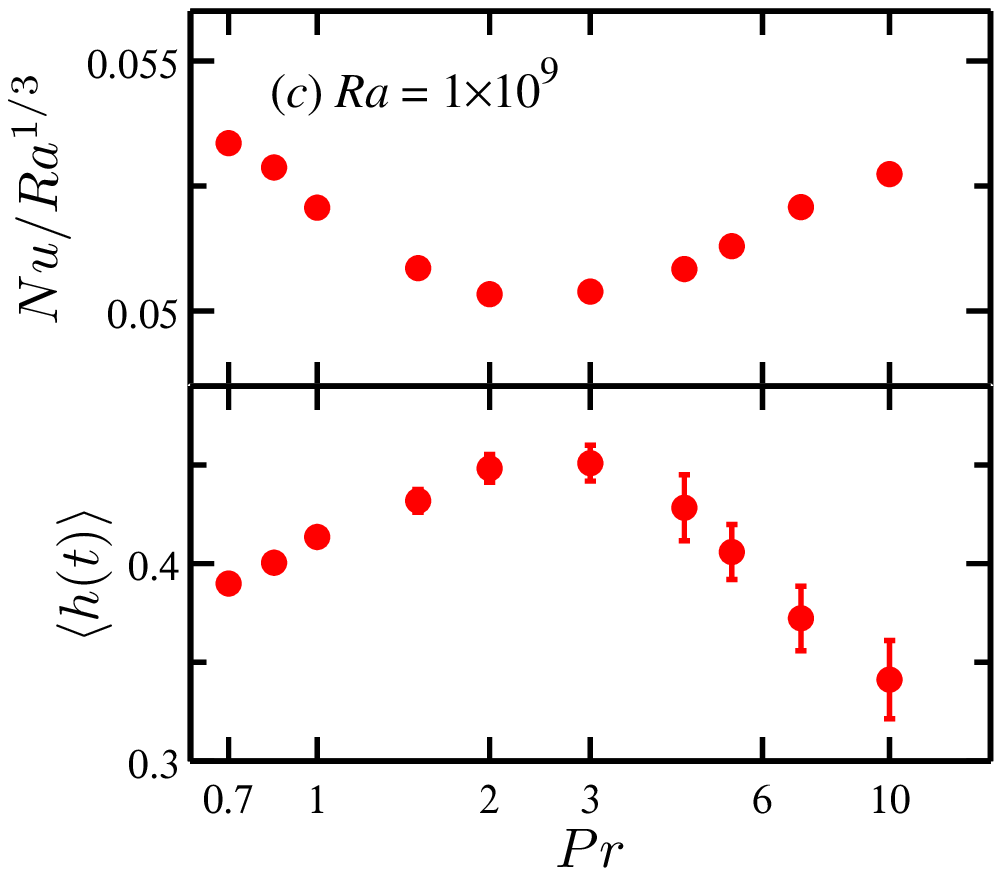}
}
\resizebox{0.495\columnwidth}{!}{%
  \includegraphics{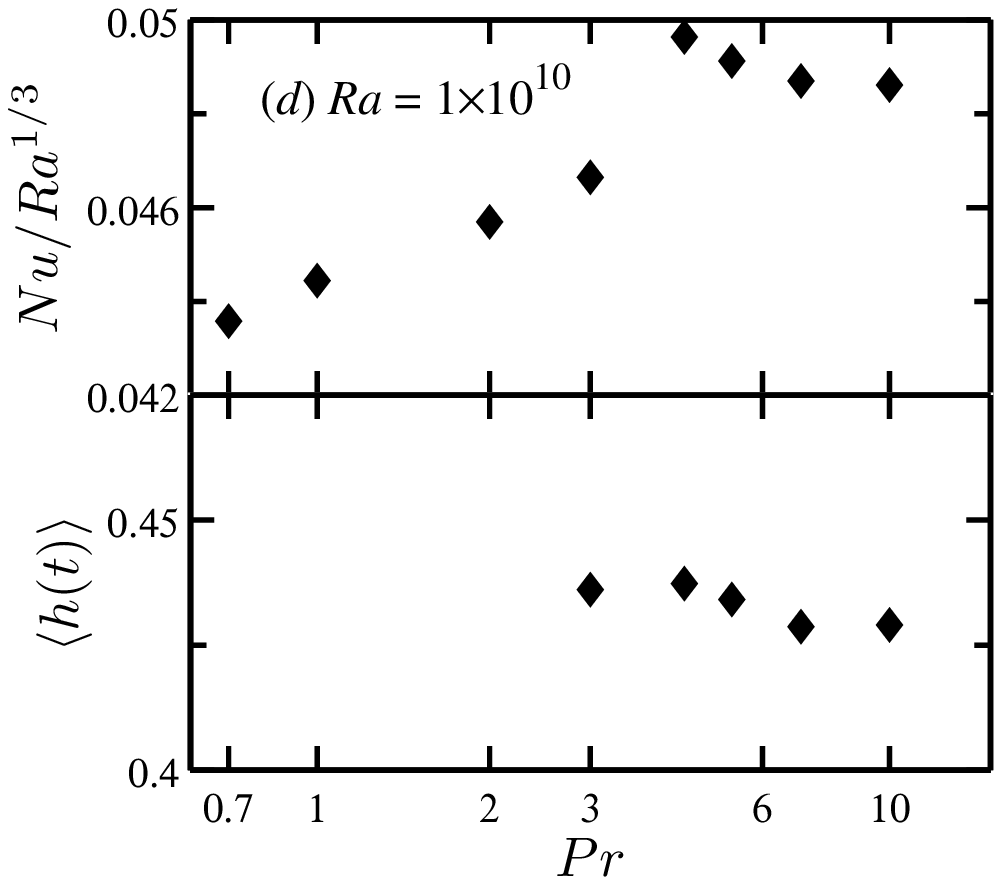}
}
\caption{(Color online). $Pr$-dependence of the compensated $Nu/Ra^{1/3}$ and the time-averaged heights $\langle h(t)\rangle$ of the lower corner flows for $Ra=1\times10^8$ (\emph{a}), $3\times10^8$ (\emph{b}), $1\times10^9$ (\emph{c}), and $1\times10^{10}$ (\emph{d}) in semilog plots. The error bars in the plots of $\langle h(t)\rangle$ indicate the standard deviations of $h(t)$. Because of the flow reversals \cite[]{xgl2010prl, chandra2011pre, chandra2013prl}, the error bars are much more pronounced for medium $Pr$ at $Ra=10^8$ in (\emph{a}) and at $Ra=3\times10^8$ in (\emph{b}). For $Ra=1\times10^{10}$ it is hard to identify the corner-flow rolls when $Pr\leq2$, see the text for the detail.} \label{fig:fig4}
\end{center}
\end{figure}

\begin{figure}
\begin{center}
\resizebox{0.495\columnwidth}{!}{%
  \includegraphics{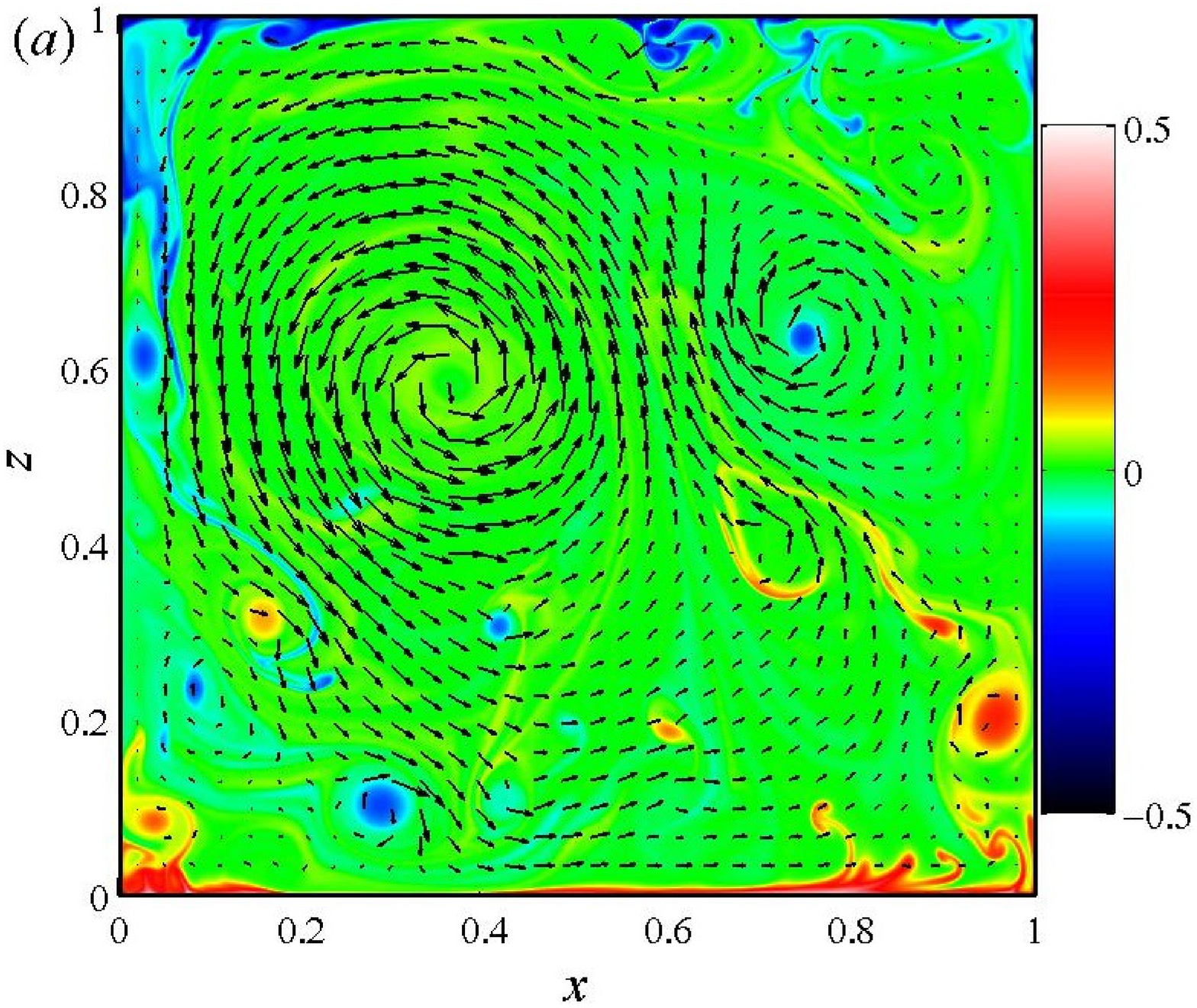}
}
\resizebox{0.495\columnwidth}{!}{%
  \includegraphics{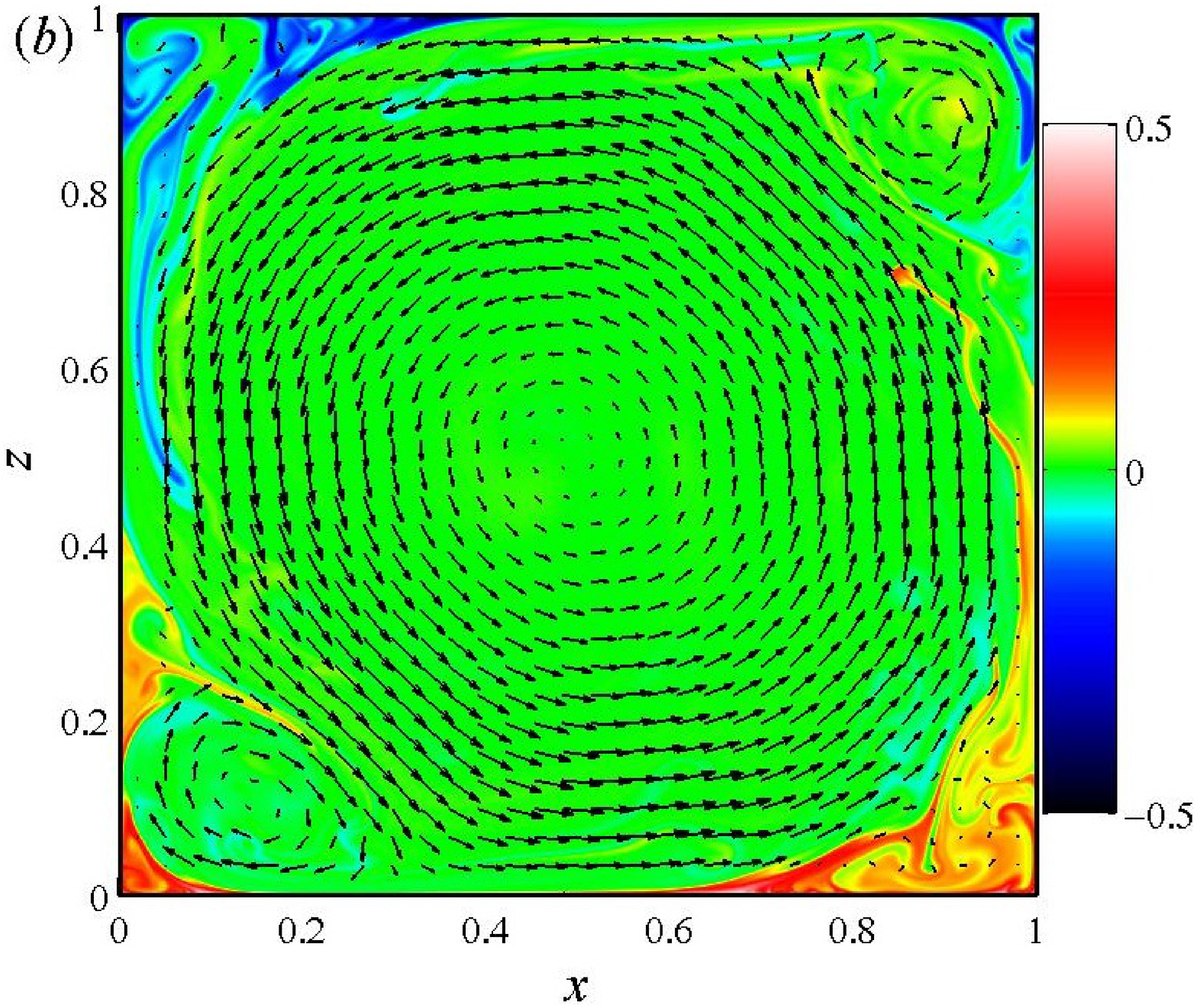}
}
\caption{(Color online). Typical snapshots of the instantaneous temperature (color) and velocity (arrows) fields for $Ra=1\times10^{10}$ and for $Pr=0.7$ (\emph{a}) and 4.38 (\emph{b}). It is seen that for low $Pr$ the corner-flow rolls can not be identified, while for large $Pr$ two larger clockwise rotating rolls at the lower-left and upper-right corners are well established.} \label{fig:fig5}
\end{center}
\end{figure}

To quantify the strength of the corner flow, we follow the idea of \cite{xgl2010prl} and adopt the height $h(t)$ of the corner flow, which can be measured by identifying the position of the steepest gradient of $\theta(z)$ at the respective sidewall. Figures \ref{fig:fig4}(\emph{a})-(\emph{c}) show the $Pr$-dependence of the time-averaged heights $\langle h(t)\rangle$ of the lower corner flows for $Ra=1\times10^8$, $3\times10^8$, and $1\times10^9$, respectively. In the figures, we also plot the compensated $Nu/Ra^{1/3}$ for comparison. It is seen that $\langle h(t)\rangle$ is highly anti-correlated with $Nu$ for $Ra$ between $10^8$ and $10^9$, i.e. there is an optimal $Pr\approx2\sim3$ for which $\langle h(t)\rangle$ is maximal. This again suggests that for moderate $Ra$ stronger corner flows restrain the global heat flux of the system via the corner-LSC competitions. Figure \ref{fig:fig4}(\emph{d}) shows $\langle h(t)\rangle$ and $Nu/Ra^{1/3}$ as functions of $Pr$ for $Ra=1\times10^{10}$. We find that when $Pr\leq2$ the corner-flow rolls are not stable and would detach from the corners (see figure \ref{fig:fig5}) and hence it is hard to identify the corner-flow rolls and quantify their heights for these flow organizations. For $Pr\geq3$, $\langle h(t)\rangle$ and $Nu$ are found to be uncorrelated, which may be due to that for high $Ra$ thermal mixing is enhanced by the increased turbulent intensity and hence the negative $Nu_L$ generated by the corner-LSC competitions is not so pronounced as that in moderate-$Ra$ cases. Taken together, these results suggest that the corner-flow rolls do not play an important role in the processes of turbulent heat transport for high $Ra$.

\section{Conclusion}

In conclusion, our high-resolution numerical measurements of $Nu$ in 2D RB convection over the parameter ranges $10^8\leq Ra\leq10^{10}$ and $0.7\leq Pr\leq10$ show that $Nu(Pr)$ minimizes at $Pr\thickapprox2\sim3$ for moderate $Ra$. This is dramatically different from the observations in 3D cylindrical cells, where the opposite situation is found, i.e. $Nu(Pr)$ maximizes near $Pr\thickapprox3$ for not too large $Ra$. We find that such a difference is attributed to the counter-gradient local heat flux generated by the competitions between the corner-flow rolls and the LSC, the effects of which are more pronounced in 2D geometry. What we want to emphasize is that although the present analysis is performed in 2D geometry, counter-gradient local heat transport is ubiquitous and it can also be found in 3D turbulent RB system \cite[]{shang2003prl,pinton2007prl}. However, due to the fluid motion in the third dimensionality, the effects of counter-gradient transport in 3D cases are not strong enough to reverse the global heat flux and make it go through a minimum as observed in 2D situations. The effects of counter-gradient heat transfer on the dynamic and global-heat-transfer processes in 3D geometry will be the objectives of our future studies.


\begin{acknowledgments}
This work was supported by Natural Science Foundation of China under Grant Nos. 11222222, 11332006, 11272196, 11202122, and 11161160554, Innovation Program of Shanghai Municipal Education Commission under Grant No. 13YZ008, and Program for New Century Excellent Talents in University under Grant No. NCET-13-0. Q. Zhou wishes to acknowledge support given to him from the organization department of the CPC central committee through `Young Talents Support Program'.
\end{acknowledgments}


\end{document}